\documentclass[12pt,a4paper]{article}
\usepackage{amssymb}
\usepackage{graphicx, overcite}
\graphicspath{ {./fig/} }

\begin{document}

\title{Non-linear elasticity effects and stratification in brushes of branched polyelectrolytes.}
\author{Inna O. Lebedeva$^{1,2}$, Oleg V.Shavykin$^{3}$, \\
Igor M.Neelov$^{3}$, Ekaterina B. Zhulina$^{3,4}$, Frans A.M.Leermakers$^5$,\\ 
Oleg V.Borisov$^{1,3,4}$ \\
$^{1}$Institut des Sciences Analytiques et de Physico-Chimie pour \\
l'Environnement et les Mat\'eriaux, UMR 5254 CNRS UPPA,\\
64053 Pau, France \\
$^{2}$Peter the Great St.Petersburg State Polytechnic University \\
195251, St.Petersburg, Russia \\
$^{3}$St.Petersburg National University of Informational Technologies,\\
Mechanics and Optics, 197101 St.Petersburg, Russia\\
$^{4}$Institute of Macromolecular Compounds \\
of the Russian Academy of Sciences, 199004 St.Petersburg, Russia \\
$^{5}$Physical Chemistry and Soft Matter, Wageningen University,\\
6703 HB Wageningen, The Netherlands}
\date{November 6, 2019}
\maketitle

\begin{center}
email: oleg.borisov@univ-pau.fr
\end{center}

\begin{abstract}
Brushes formed by arm-tethered starlike polyelectrolytes may exhibit
internal segregation into weakly and strongly extended populations
(stratified two-layer structure) when strong ionic intermolecular repulsions
induce stretching of the tethers up to the limit of their extensibility. 
We propose an approximate Poisson-Boltzmann theory for analysis of the
structure of the stratified brush and compare it with results of numerical
self-consistent field modelling. Both analytical and numerical models point
to formation of a narrow cloud of counterions (internal double electrical
layer) localized inside stratified brush at the boundary between the layers.
\end{abstract}

\newpage

\section{Introduction}

Modification of solid-liquid interface by layers of anchored macromolecules
("polymer brushes") enables tuning interaction and friction forces between
surfaces providing thereby a robust approach to control the aggregative
stability of colloidal dispersions \cite{Napper,Isrch,book} and boundary
lubrication \cite{JKlein_EKumacheva1994,Schorr2003,Raviv_JKlein2002,
Spencer2008,Armes_JKlein2009}.

The use of tethered ionically charged macromolecules (polyelectrolytes) in
aqueous medium makes it possible to exploit long-range electrostatic
interactions 
that are easily tunable by varying the ionic strength and (in the case of
weak polyelectrolytes) pH of the solution. 
Brushes of charged macromolecules are also exploited by nature. For example,
thick extracellular layers of polysaccharides (glycocalyx) decorating
bacterial surfaces mediate inter-cell interaction and adhesion \cite%
{Camesano2002,Abu-Lail2003,Button2012,Vu2009}. Some of these polysaccharides
are branched (have tree-like architecture). The branched
architecture (topology) of macromolecules can be thereby considered as one
of the design parameters in technological and biomedical applications of
polymer brushes \cite{Paez2012,Gillich2011,Gillich2013,Schull2013,Borisov2014,Leermakers2017}.


While structure of interfacial layers formed by linear chain
polyelectrolytes is comprehended on the basis of existing theories and
supporting them experimental data \cite{Ballauff,Minko,Toomey,Ruhe2004}, our
knowledge about interplay between branching of the brush-forming
macromolecules and ionic interactions is still incomplete.

In particular, ionic intermolecular interactions operating in
polyelectrolyte brushes can cause strong stretching of the brush-forming
chains. As a result, the most stretched (proximal to the grafting surface)
segments approach the limit of extensibility when the fraction of charged
monomer units in polyelectrolyte chains is sufficiently large or/and ionic
strength of the solution is low \cite%
{Misra1991,BA2000,BVA,Netz2003,Netz2004,Lebedeva2017}. Even in the brushes
formed by linear polyelectrolyte chains 
the distribution of elastic tension along the contour of the chains is
essentially non-uniform and decreases as a function of the distance from the
grafting surface. However, as demonstrated in ref \cite{Lebedeva2017}, the
account of finite chain extensibility within the self-consistent field
Poisson-Boltzmann approach does not lead to qualitatively different
predictions concerning the brush structure as compared to the theory built
up using Gaussian (linear) elasticity approximation \cite{BZ2,BZKW}.

The situation becomes more dramatic for brushes formed by
dendritically-branched (tree-like) polyelectrolytes: Here due to increasing
number of spacers/branches in higher generations the distribution of elastic
tension is strongly non-uiform and sharply decreases as a function of the
generation ranking number. The most strongly stretched is the stem by which
the dendron is linked to the surface, whereas only minor stretching is
expected for the free branches. As demonstrated in refs \cite%
{Polotsky10-1,Polotsky12-1,Zhulina2014,Borisov2014}, even in non-ionic dendron
brushes governed by excluded volume intermolecular interactions, the stem
can easily approach the limit of extensibility that leads to a specific for
the dendron brushes effect of stratification. This effect is most pronounced
in brushes made up by dendrons of the first generation, i.e., arm-tethered
starlike polymers that segregate in two populations with strongly and
moderately extended stems \cite{Polotsky12-1,Zhulina2014}.

The aim of the present paper is to study the effects resulting from finite
extensibility (non-linear elasticity) in brushes formed by ionically charged
first generation dendrons (arm-tethered polyelectrolyte stars). In
particular, we examine the equilibrium structure of a stratified brush with
focus on the distributions of polymer density, the end-points of free arms,
the branching points, and local charge density. For that we propose an
approximate Poisson-Boltzmann analytical approach and complement it 
with the numerical self-consistent field calculations.

\section{Brush of arm-tethered polyelectrolyte stars}


Consider a planar brush composed of stars with $q+1$ branches (arms), with
one branch (stem) attached to the surface by the terminal segment, and $q$
free branches, \textbf{Figure 1}. Each branch has degree of polymerization $n
$ and fraction of permanently (positively) charged monomer units $\alpha $.
Total number of monomer units in the macromolecule is $N=n(1+q)$, the total
charge of a star is $Q(q)=\alpha (q+1)n$. All the branches are assumed to be
intrinsically flexible with the Kuhn segment length on the order of monomer
size $a\simeq l_{B}$, where $l_{B}=e^{2}/\varepsilon k_{B}T$ is the Bjerrum
length. Macromolecules are tethered with an area $s$ per star (or,
equivalently, with grafting density $\sigma =a^{2}/s$). The brush is
immersed in the solution containing monovalent salt with respective bulk
concentrations of co- and counterions $c_{+}=c_{-}=c_{s}$ that specify the
Debye screening length as $\kappa ^{-1}=(8\pi l_{B}c_{s})^{-1/2}$ 

\begin{figure}[ht]
\centering{\includegraphics[scale=0.5]{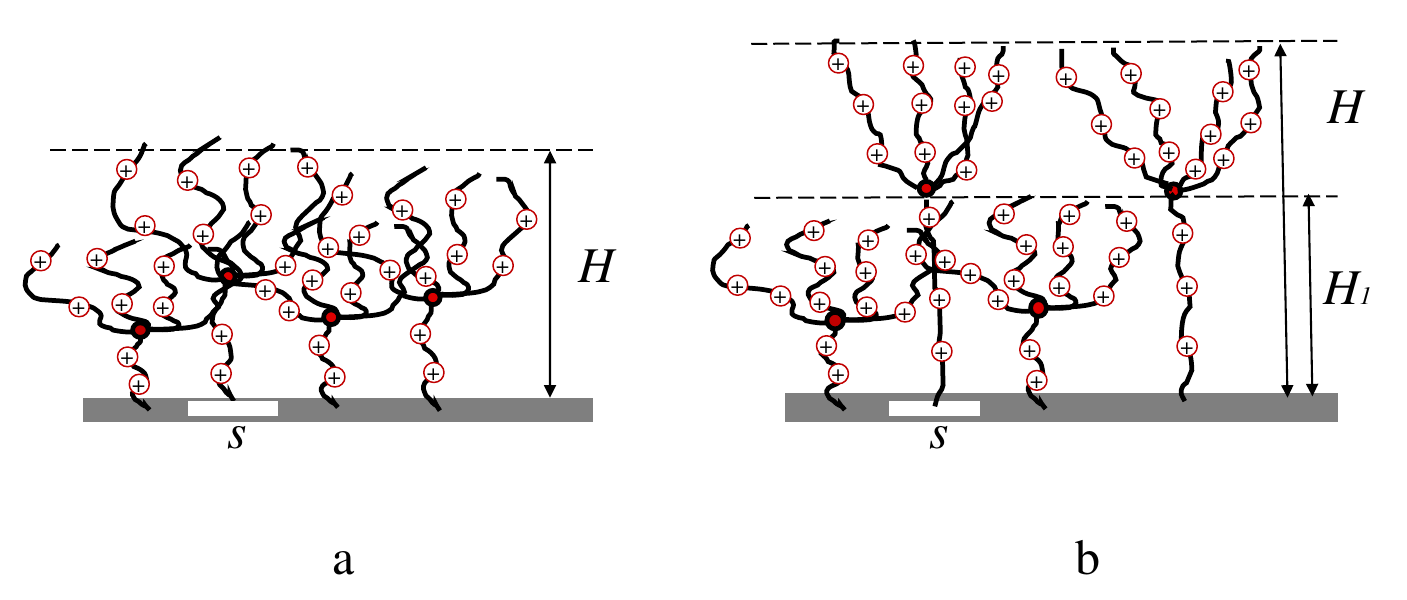}}   
\caption{ Schematics of the brush of arm-tethered polyelectrolyte stars in
linear elasticity regime (a) and in stratified regime (b)}
\label{Fig1}
\end{figure}


The electrostatic interactions between all charged species (ionized monomer
units and mobile ions) are described within the accuracy of non-linear
Poisson-Boltzmann framework, that is, through the self-consistent
electrostatic potential $\Psi (z)$ which is a function of the distance $z$
from the grafting surface.


In order to develop an analytical theory for the brush of polyelectrolyte
stars the Poisson-Boltzmann approach is coupled to strong stretching (SS)
approximation \cite{Semenov} that assumes significant extension of stems and
free branches of tethered macromolecules with respect to their Gaussian
dimensions.



We further assume that the self-consistent molecular potential $U(z)$ acting
in the brush 
is dominated by ionic interactions. Then it coincides with the electrostatic
energy $\alpha e\Psi(z)$ per monomer: 
\begin{equation}
\frac{U(z)}{k_{B}T}\approx \frac{\alpha e\Psi _{in}(z)}{k_{B}T}=\alpha
\psi_{in}(z)  \label{Psi}
\end{equation}
where $\psi _{in}(z)=e\Psi_{in}(z)/k_{B}T$ is the reduced (dimensionless)
electrostatic potential at distance $z$ from the surface measured in $k_{B}T$
units. The explicit form of the molecular potential depends on degree of
extension of linear segments of the branched macromolecules.

\section{Gaussian (linear) elasticity regime}

In the case when all the linear segments are considerably extended with
respect to their ideal dimensions but far below the contour length and thus
exhibit Gaussian conformational elasticity (linear response regime), the
molecular potential for the brush formed by regular dendritically-branched
macromolecules attached to the surface through the focal points has the
simple quadratic form 
\begin{equation}
\frac{U(z)}{k_{B}T}=\frac{3}{2a^{2}}k^{2}(H^{2}-z^{2})  \label{U}
\end{equation}%
where $H$ is the brush thickness and $k$ is so-called topological
coefficient which is fully determined by the topology of the brush-forming
macromolecules but is remarkably independent of the strength and ionic or
non-ionic character of interactions in the system \cite%
{Pickett2001,Borisov2014}. For brushes of arm-tethered stars (the first
generation dendrons) the topological coefficient $k$ was calculated in ref 
\cite{Polotsky12-1} using the condition of elastic force balance in the
branching point and equals 
\begin{equation}
k=n^{-1}\arctan (\frac{1}{\sqrt{q}})  \label{k}
\end{equation}

The structure of star polyelectrolyte brush in the linear elasticity regime
was studied elsewhere \cite{PE_starbrush_linear} and here we only briefly
summarize the results.

By combining eqs \ref{Psi} and \ref{U} we find the expression for
electrostatic potential in the brush as 
\begin{equation}
\psi_{in}(z) = \frac{H^{2}-z^{2}}{H_{0}^2(q)}  \label{psix}
\end{equation}
where the characteristic length 
\begin{equation}
H_0(q)= \sqrt{\frac{2}{3}}\frac{a\alpha ^{1/2}n}{\arctan(1/\sqrt{q})}
\label{H0}
\end{equation}
depends on the number of arms in starlike polyions. We introduce also the
corresponding length for the brush of linear polyions of length $n$, 
\begin{equation}
H_0 = \sqrt{\frac{8}{3\pi^2}}\alpha ^{1/2}na  \label{H0_lin}
\end{equation}
which formally follows from eq \ref{H0} at $q=0$. Then 
\[
\frac{H_0}{H_{0}(q)}=\frac{2}{\pi}\arctan\frac{1}{\sqrt{q}} 
\]

By applying the Poisson equation

\begin{equation}
\frac{d^{2}\psi _{in}(z)}{dz^{2}}=-4\pi l_{B}\rho (z)  \label{P}
\end{equation}
together with eq \ref{psix}, one finds the net number charge density $\rho(x)
$ inside the brush as 
\begin{equation}
\rho (z)=\alpha c(z)+c_{+}(z)-c_{-}(z)=\frac{1}{2\pi l_{B}H_{0}^{2}(q)}
\label{ro}
\end{equation}
The cumulative (residual) charge of the brush layer of thickness  $z$ is 
\begin{equation}
\tilde Q(z)=\int_{0}^{z}\rho (z^{\prime})dz^{\prime}=\frac{z}{2\pi
l_{B}H_{0}^{2}(q)}  \label{Qz}
\end{equation}
which increases linearly as a function of the layer thickness $z$. The
residual (uncompensated) charges $\tilde Q$ per unit area of the brush, 
\begin{equation}
\tilde Q(H)=\int_{0}^{H}\rho (z')dz'=\frac{H}{2\pi l_{B}H_{0}^{2}(q)}
\label{Qf}
\end{equation}

The latter determines the Gouy-Chapman length 
\[
\widetilde{\Lambda }(q)=\frac{1}{2\pi l_{B}\tilde Q(H)}=\frac{H_{0}^{2}(q)}{H%
} 
\]
associated with the electrostatic potential $\psi _{out}(z)$ and
distribution of co- and counterions outside the brush, i.e. at $z\geq H$%
.

As one can see from eq \ref{ro}, within Gaussian (linear) elasticity
approximation, the net charge density inside the brush is constant
(independent of the distance $z$ from the surface).

The mobile ions are distributed according to the Boltzmann law as 
\begin{equation}
c_{\pm }(z)=c_{\pm }(H)\exp [\mp \psi (z)]  \label{cc}
\end{equation}
with%
\[
\psi (z)=\left\{ 
\begin{array}{cc}
\psi _{in}(z) & 0\leq z \leq H \\ 
\psi _{out}(z) & z \geq H%
\end{array}
\right. 
\]

where\cite{BZKW}

\begin{equation}
\psi _{out}(z)=2\ln \left[ \frac{(\kappa \tilde{\Lambda}(q)+\sqrt{(\kappa 
\tilde{\Lambda}(q))^{2}+1}-1)+(\kappa \tilde{\Lambda}(q)-\sqrt{(\kappa 
\tilde{\Lambda}(q))^{2}+1}+1)e^{-\kappa (z-H)}}{(\kappa \tilde{\Lambda}(q)+%
\sqrt{(\kappa \tilde{\Lambda}(q))^{2}+1}-1)-(\kappa \tilde{\Lambda}(q)-\sqrt{%
(\kappa \tilde{\Lambda}(q))^{2}+1}+1)e^{-\kappa (z-H)}}\right]
\label{Psiout}
\end{equation}

and 
\[
c_{\pm }(H)= 
c_{s}\left( \frac{\sqrt{(\kappa \widetilde{\Lambda}(q) )^{2}+1}-1}{\kappa 
\widetilde{\Lambda}(q) }\right) ^{\pm 2} 
\]%
\[
=c_{s}\left( \frac{\sqrt{(\kappa \widetilde{\Lambda}(q) )^{2}+1}\mp 1}{%
\kappa \widetilde{\Lambda}(q) }\right) ^{2} 
\]

Hence, within the linear elasticity regime, the concentration of counterions
inside the brush smoothly decreases with the distance from the grafting
surface as a Gaussian function of $z$, whereas outside the brush it decays
with the characteristic length $\sim \mbox{min}\{\kappa ,\tilde{\Lambda}(q)\}
$ which latter coincides with the thickness of the counterion could,
neutralizing the residual charge of the brush.

The density profile of charged monomer units $\alpha c(x)$ is then
determined from eq \ref{ro} as 
\[
\alpha c(z)=\frac{1}{2\pi l_{B}H_{0}^{2}(q)}+c_{-}(z)-c_{+}(z)=\frac{1}{2\pi
l_{B}H_{0}^{2}(q)}+ 
\]
\begin{equation}
+c_{s}\left( \frac{\sqrt{(\kappa \widetilde{\Lambda}(q) )^{2}+1}+1}{\kappa 
\widetilde{\Lambda}(q) }\right) ^{2}\exp \left[ \frac{(H^{2}-z^{2})}{%
H_{0}(q)^{2}}\right] -c_{s}\left( \frac{\sqrt{(\kappa \widetilde{\Lambda}(q)
)^{2}+1}-1}{\kappa \widetilde{\Lambda}(q) }\right) ^{2}\exp \left[ -\frac{%
(H^{2}-z^{2})}{H_{0}(q)^{2}}\right]  \label{ac}
\end{equation}

By integrating the polymer density profile, 
\[
\int_{0}^{H}c(z)dz=\frac{(q+1)n}{s}
\]%
one gets the equation for reduced brush thickness $h=H/H_{0}(q)$ as a
function of two dimensionless parameters, $\kappa H_{0}(q)$ and 
\[
\zeta =2\pi l_{B}\alpha (q+1)nH_{0}(q)/s\sim n^{2}(q+1)\arctan ^{-1}\frac{1}{%
\sqrt{q}}
\]%
as

\[
\zeta =h+\left( \sqrt{(\frac{\kappa H_{0}(q)}{2})^2 +h^{2}/4}+h/2\right)
^{2}\int_{0}^{h}\exp (h^{2}-\xi ^{2})d\xi 
\]
\begin{equation}
-\left( \sqrt{(\frac{\kappa H_{0}(q)}{2})^2 + h^{2}/4}-h/2\right)^{2}%
\int_{0}^{h}\exp [-(h^{2}-\xi ^{2})]d\xi  \label{hf}
\end{equation}

with asymptotic solutions

\begin{equation}
h=H/H_0(q) \approx \left\{ 
\begin{array}{ll}
\zeta, & \zeta \ll \mbox{min}\{1,(\kappa H_0(q))^{-1}\} \\ 
\sqrt{\ln(2\zeta\sqrt{\pi})}, & \zeta\gg \mbox{max}\{1,(\kappa H_0(q))^{2}\}
\\ 
(3\zeta(\kappa H_0(q))^{-2}/4)^{1/3}, & (\kappa H_0(q))^{-1}\ll \zeta \ll
(\kappa H_0(q))^{2}.%
\end{array}
\right.  \label{h_PB_salt_limits}
\end{equation}

The first two lines in eq \ref{h_PB_salt_limits} describe low-salt regimes,
among them the osmotic regime (corresponding to the second line) is
experimentally most relevant: In the osmotic regime $\tilde{Q}(H)\ll Q$,
that is, the residual charge of the brush is much smaller than its bare
charge. In the osmotic regime the thickness of the brush grows only weakly as a function
of the number of arms  as 
$$H\sim \arctan ^{-1}\frac{1}{\sqrt{q}}%
\sqrt{\ln ((q+1)\arctan ^{-1}\frac{1}{\sqrt{q}})}
$$ 
As a result, the
concentration of the counterions entrapped inside the brush rapidly
increases as a function of the number of arms in the star. Therefore, larger salt
concentration is required for triggering contraction of the brush caused by
salt-induced screening of electrostatic interactions as described by the
third line in eq \ref{h_PB_salt_limits}.

\section{Non-linear elasticity regime}

The limit of the linear elasticity regime corresponds to stretching of the
(most extended) stems up to their contour length. This can be readily
achieved upon an increase in the fraction $\alpha$ of charged monomer units
in the stars.

For describing the nonlinear elasticity regime for the brush of tethered starlike
polyions we follow the route outlined in ref \cite{Zhulina2014} for analysis
of the structure of brushes formed by arm-tethered neutral (non-ionic) stars.

As an essential prerequisite of the theory, we take advantage of the known
molecular potential $U(z)$ in the brush of linear chains of $n$ monomers
with finite extensibility on bcc (body centered cubic) lattice \cite{BA2000} 
\begin{equation}
\frac{U(z)}{k_{B}T}=3\ln \cos (\frac{\pi z}{2an})+const  \label{UL}
\end{equation}%
In considering the brush of strongly extended starlike polymers we adopt the
approximate analytical two-layer model in which the brush consists of two
(lower and upper) layers, and the stars are splitted into two respective
populations.

The lower (proximal to the grafting surface) layer of thickness $H_{1}$
contains (i) fraction $1-\beta $ of stars which are relatively weakly
stretched and completely embedded into the lower layer and (ii) strongly
stretched stems of the fraction $\beta$ of stars whose free branched compose
the upper layer. The upper layer of thickness $H-H_1$ is formed by free branched of the latter population of stars. Their branching
points are all localized at $z=H_1$. Hence, the upper layer is equivalent to
the brush of linear chains of length $n$ with the grafting density $q\beta
a^2/s$

For describing the molecular potential in the lower layer $U_{1}(z)$, we
adopt the same form of $U_{1}(z)$ as in eq \ref{UL} but replace $%
k_{lin}=\pi/2n$ by the topological coefficient for the brush of stars 
\[
k=k_{star}(q)=n^{-1}\arctan(1/\sqrt{q}) 
\]
that is 
\begin{equation}
\frac{U_{1}(z)}{k_BT}\approx \lambda _{1}+3\ln \cos (\frac{kz}{a})
\label{U1}
\end{equation}

In the upper layer, which is equivalent to the brush of linear chains of
length $n$ (with $k=k_{lin}=\pi/2n$) the molecular potential is given by 
\begin{equation}
\frac{U_{2}(z)}{k_BT}= 3\ln \frac {\cos \frac{\pi (z-H_1)}{2an}}{\cos \frac{%
\pi (H-H_1)}{2an}}  \label{U2}
\end{equation}
that ensures vanishing of $U_2(z)$ at $z=H$.

The condition of continuity of the molecular potential at $z=H_1$, that is $%
U_1(z=H_1)=U_2(z=H_1)$ enables us to determine the constant $\lambda_1$ in
eq \ref{U1} as 
\[
\lambda_1=-3\ln \biggl( \cos \frac{kH_1}{a}\cdot \cos \frac{\pi(H-H_1)}{2an}%
\biggr)
\]

The reduced self-consistent electrostatic potential in the brush $\psi_{in}(z)=e%
\Psi(z)/k_BT=U(z)/\alpha k_BT$ is thus given by

\begin{equation}
\psi_{in}(z)= \frac{3}{\alpha}\left\{ 
\begin{array}{ll}
\ln \frac{\cos\frac{kz}{a}}{\cos \frac{kH_1}{a}\cos \frac{\pi(H-H_1)}{2na}} ,
& 0\leq z\leq H_1 \\ 
\ln \frac{\cos \frac{\pi(z-H_1)}{2na}}{\cos\frac{\pi(H-H_1)}{2na}}, & 
H_1\leq z \leq H%
\end{array}
\right.  \label{psi_bilayer}
\end{equation}

Electrostatic potential $\psi_{in}(z)$ in eq \ref{psi_bilayer} exhibits two 
distinct features: (\textit{i}) it is continuous at the boundary between 
the layers, at $z=H_{1}$; (\textit{ii}) the first derivative $d\psi_{in}/dz$
which  is proportional to the strength of electrostatic field, exhibits a
jump at $ z=H_{1}$ from a finite value at $z=H_{1}-0$\ to zero at $z=H_{1}+0$%
.

The net charge density $\rho(z)$ inside both layers can be found from eq \ref%
{psi_bilayer} by using the Poisson equation, eq \ref{P}: 
\begin{equation}
\rho(z)= \frac{1}{2\pi l_B H_{0}^{2} }\left\{ 
\begin{array}{ll}
\frac{H_{0}^{2}}{H_{0}^{2}(q)}\sec^2\frac{kz}{a} , & 0\leq z\leq H_1 \\ 
\sec^2 \frac{\pi(z-H_1)}{2na} , & H_1\leq z \leq H%
\end{array}
\right.  \label{rho_z_1_2}
\end{equation}

The residual charge in the brush (per unit area) within proximal layer (layer 1) is given by
\begin{equation}
\tilde{Q}_{1}=\int_{0}^{H_{1}}\rho (z)dz=\frac{1}{2\pi l_{B}H_{0}^{2}}\biggl(%
\frac{H_{0}^{2}}{H_{0}^{2}(q)}\frac{a}{k}\tan \frac{kH_{1}}{a}\biggr)
\label{Q1}
\end{equation}

Since according to eq \ref{psi_bilayer} the strength of electrostatic field $%
d\psi_{in}/dz$ at $z=H_1+0$ is zero, the residual charge $\tilde Q_1$ of the
proximal layer given by eq \ref{Q1} should be neutralized  by infinitely
thin cloud of counterions localized at $z\approx H_1$. That is, $%
c_{-}(z)=\tilde Q_1\delta(z-H_1)$ at $z\approx H_1$ where $\delta(x)$ is the
Dirac delta-function.

The residual charge in the brush (per unit area) in the peripheral layer (layer 2) is given by
\begin{equation}
\tilde{Q}_{2}=\int_{H_{1}}^{H}\rho (z)dz=\frac{1}{2\pi l_{B}H_{0}^{2}}\biggl(%
\frac{2na}{\pi }\tan \frac{\pi (H-H_{1})}{2na}\biggr)
\end{equation}

and it is related to the Gouy-Chapman length outside the brush 
\begin{equation}
\tilde{\Lambda}(q)=\frac{1}{2\pi l_{B}\tilde{Q}_{2}}=\frac{H_{0}^{2}}{a}%
\biggl(\frac{2na}{\pi }\tan \frac{\pi (H-H_{1})}{2na}\biggr)^{-1}
\end{equation}%
which controls the distribution of electrostatic field and concentration
profile of ions outside the brush.

As a particular case, we consider stratified brush of starlike
polyelectrolytes in a salt-free solution which contains (monovalent)
counterions only.

Mobile counterions outside of the brush are distributed similarly to that
from a uniformly charged surface with the surface charge density $\widetilde{%
Q}$ in contact with the salt-free solution. The concentration of ions at $x=H
$ is thus given by, 
\begin{equation}
c_{-}(H)=\frac{1}{2\pi l_{B}\widetilde{\Lambda }^{2}(q)}
\end{equation}
that is 
\begin{equation}
c_{-}(H)=\frac{a^2}{2\pi l_{B} H_{0}^4}\biggl(\frac{H_{0}^2}{H_{0}^{2}(q)}%
\frac{a}{k}\tan\frac{kH_1}{a}+\frac{2na}{\pi}\tan\frac{\pi(H-H_1)}{2na}%
\biggr)^{2}
\end{equation}

In the proximal layer (layer 1) profile of concentration of counterions is given by 

\[
c_{1-}(z)= c_{-}(H) \exp(\psi_1(z)) = 
\]
\begin{equation}
\frac{a^2}{2\pi l_{B} H_{0}^4}\biggl(\frac{H_{0}^2}{H_{0}^{2}(q)}\frac{a}{k}%
\tan\frac{kH_1}{a}+\frac{2na}{\pi}\tan\frac{\pi(H-H_1)}{2na}\biggr)^{2} %
\biggl(\frac{\cos \frac{kz}{a}}{\cos \frac{kH_1}{a} \cos\frac{\pi(H-H_1)}{2na%
}}\biggr)^{\frac{3}{\alpha}}
\end{equation}
Then polymer density in the lower layer $\alpha c_1(z)=\rho_1(z)+ c_{1-}(z)$
is given by 
\[
\alpha c_{1}(z)=\frac{1}{2\pi l_B H_{0}^{2}(q)}\sec^2\frac{kz}{a}+ 
\]
\begin{equation}
\frac{a^2}{2\pi l_{B} H_{0}^4}\biggl(\frac{H_{0}^2}{H_{0}^{2}(q)}\frac{a}{k}%
\tan\frac{kH_1}{a}+\frac{2na}{\pi}\tan\frac{\pi(H-H_1)}{2na}\biggr)^{2} %
\biggl(\frac{\cos \frac{kz}{a}}{\cos \frac{kH_1}{a} \cos\frac{\pi(H-H_1)}{2na%
}}\biggr)^{\frac{3}{\alpha}}
\end{equation}

In the peripheral layer (layer 2)

\[
c_{2-}(z)= c_{-}(H) \exp(\psi_2(z)) = 
\]
\begin{equation}
\frac{a^2}{2\pi l_{B} H_{0}^4}\biggl(\frac{H_{0}^2}{H_{0}^{2}(q)}\frac{a}{k}%
\tan\frac{kH_1}{a}+\frac{2na}{\pi}\tan\frac{\pi(H-H_1)}{2na}\biggr)^{2} %
\biggl(\frac{\cos \frac{\pi(z-H_1)}{2an}}{\cos \frac{\pi(H-H_1)}{2an}}\biggr)%
^{\frac{3}{\alpha}}
\end{equation}
Then polymer density in the upper layer $\alpha c_2(z)=\rho_2(z)+ c_{2-}(z)$
is given by 
\[
\alpha c_{2}(z)=\frac{1}{2\pi l_B H_{0}^{2}(q)}\sec^2\frac{\pi(z-H_1)}{2na}
+ 
\]
\begin{equation}
\frac{a^2}{2\pi l_{B} \alpha H_{0}^4}\biggl(\frac{H_{0}^2}{H_{0}^{2}(q)}%
\frac{a}{k}\tan\frac{kH_1}{a}+\frac{2na}{\pi}\tan\frac{\pi(H-H_1)}{2na}%
\biggr)^{2} \biggl(\frac{\cos \frac{\pi(z-H_1)}{2an}}{\cos \frac{\pi(H-H_1)}{%
2an}}\biggr)^{\frac{3}{\alpha}}
\end{equation}

As we shall see in the following section, the thickness $H_{1}$ of the lower
layer equals, with a good accuracy to the length $an$ of fully extended arm
of the star (tether) whereas 
the overall brush thickness $H$ is an increasing function of the degree of
ionization $\alpha $.

\subsection{Self-consistent field numerical modelling}

In order to demonstrate appearance of the two-layer structure in a brush of
arm-tethered polyelectrolyte stars and to verify  the approximate analytical
model of a stratified brush we performed a series of calculations using
numerical Scheutjens-Fleer self-consistent field method \cite{book}.  More specifically we have studied brushes with 
progressively
increasing fraction $\alpha$ of (permanently) charged monomer units in
contact with solution comprising low concentration (volume fraction) of added salt.

\begin{figure}[ht]
\centering{\includegraphics[scale=0.3]{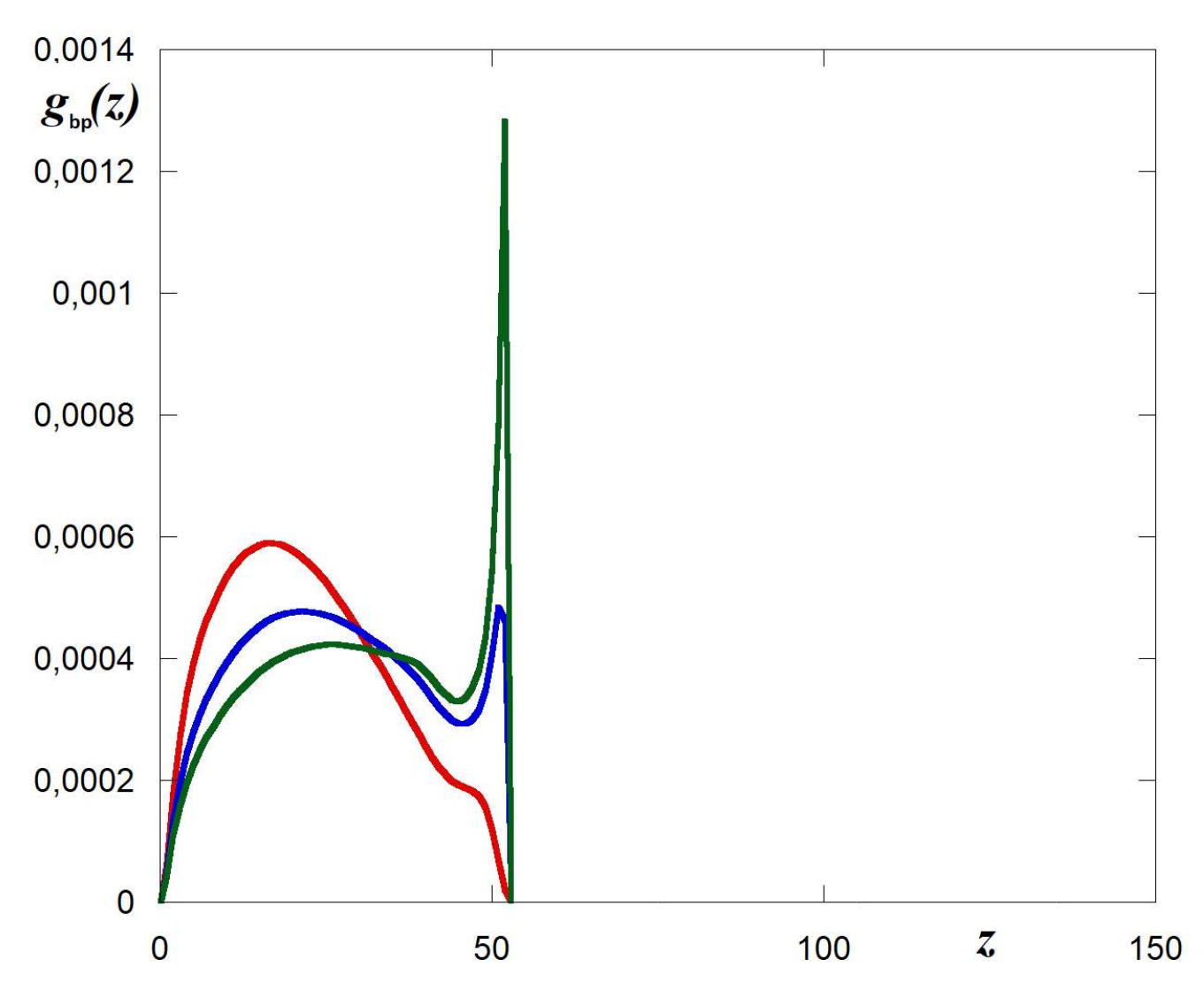}}   
\caption{ Branching points distribution in the brush of arm-tethered
starlike polyelectrolytes. Red line corresponds to $\protect\alpha=0.3$,
blue line - $\protect\alpha=0.5$, green line - $\protect\alpha%
=0.7$. Other parameters are $n=50,q=3, a^2/s=0.02$, salt volume fraction $c_s=10^{-5}$}
\label{bp}
\end{figure}

\begin{figure}[ht]
\centering{\includegraphics[scale=0.3]{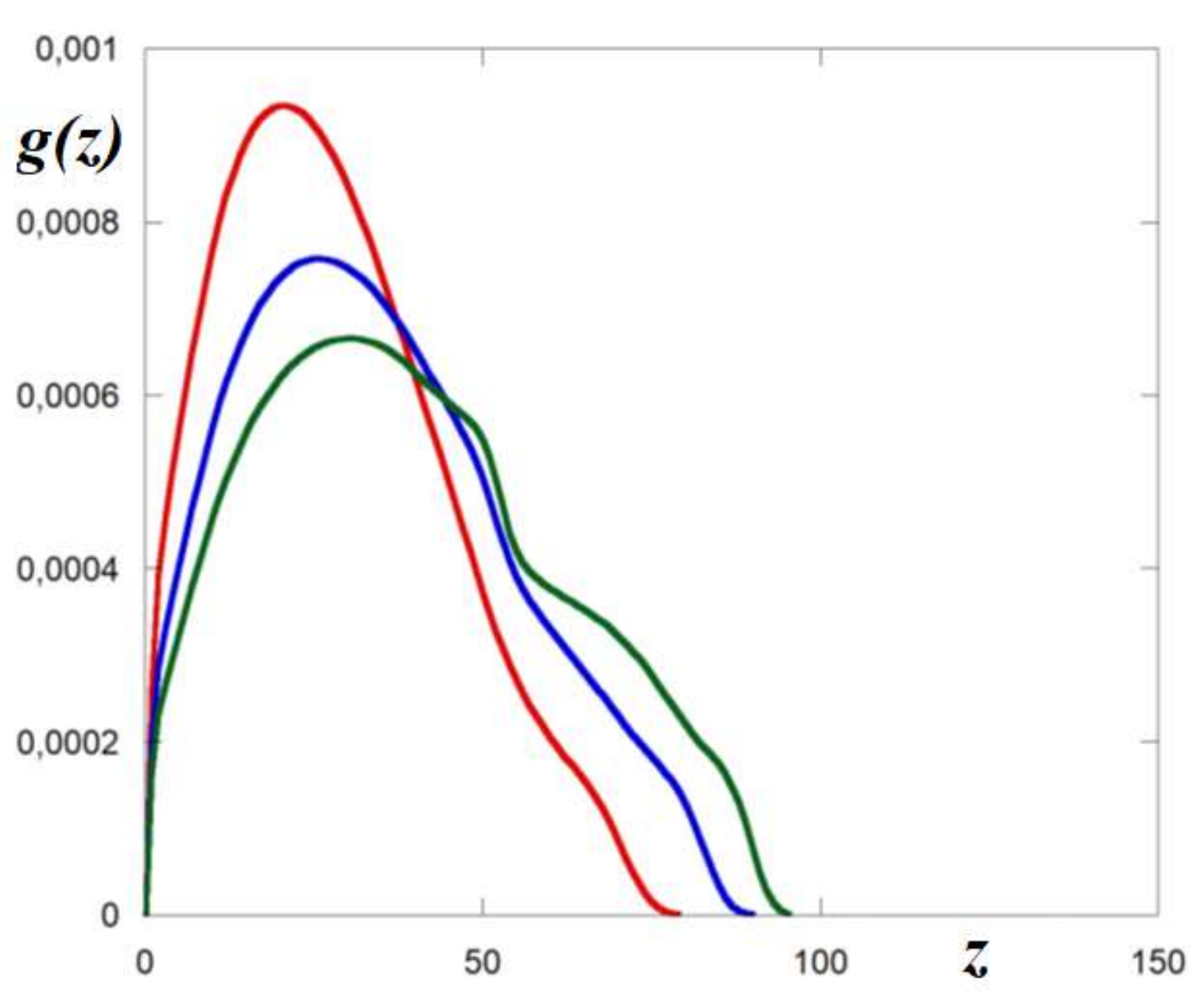}}   
\caption{ Free ends distribution in the brush of arm-tethered starlike
polyelectrolytes. Red line corresponds to $\protect\alpha=0.3$, blue line
- $\protect\alpha=0.5$, green line - $\protect\alpha=0.7$.
Other parameters are $n=50,q=3, a^2/s=0.02$, salt volume fraction $c_s=10^{-5}$.}
\label{ends}
\end{figure}

The most straightforward way to monitor how stratification appears in the
brush is to analyze the evolution of distributions of branching points and
free ends of the arms with respect to the grafting surface upon an increase
in $\alpha$. These two distributions are presented in Figures \ref{bp} and %
\ref{ends}, respectively. Each of the distributions demonstrates single
maximum with a weakly pronounced shoulder in the case of smallest fraction
of charged monomer units, $\alpha=0.3$. Since, as one can see in Figure \ref%
{bp}, at $\alpha=0.3$ only a small fraction of stems approach the limit of
extensibility ($z=50$), we anticipate that at $\alpha=0.3$ the brush is in
the transition between regimes of linear and non-linear elasticity. The
shoulders in the distributions of the end segments and branching points
indicate emerging, but not yet pronounced stratification.

An increase in $\alpha$ results in the increase in the total thickness of the
brush that corresponds to stronger stretching of the stems and branches of
the stars. Moreover, at $\alpha=0.5$ a second peak at $z\approx 50$ appears
it the distribution of branching points.

This peak corresponds to the population of stars with almost fully stretched
stems that coexist with another population with moderately stretched stems.
The latter population corresponds to localized in the inner region of the brush (at $z\approx 20$).
wide maximum in the branching points
distribution  Upon further increase in $\alpha $ the peak in the branching
points distribution at $z\approx 50$ becomes even higher that reflects an
increase in the fraction of stars with fully stretched stems (repartitioning
between weakly and strongly extended stars populations). Simultaneously the
proximal peak decreases in the magnitude and is shifter to larger values of $%
z$, i.e., the average extension of stars constituting the weakly stretched
population also increases upon an increase in $\alpha $.

The distribution of the free ends of the star arms, Figure \ref{ends},
demonstrates a similar trend: it has only one wide maximum with a weak
shoulder  at small $\alpha$, whereas at large $\alpha$ a well-pronounced
shoulder 
appears closer to the edge of the brush.

The net local charge density $\rho (z)$ and its integral $\tilde{Q}%
(z)=\int_{0}^{z}\rho (z')dz'$ are plotted as a function of $z$ in Figure \ref%
{charge}. For $\alpha =0.3$ the net charge density is fairly constant inside
the brush (cf. eq \ref{ro}), exhibits a peak at the edge of the brush $%
z\approx H$ due to loss of stretching at the ends of the free arms (not
accounts for within SS-SCF formalism) and then a deep and wide minimum
corresponding to the cloud of counterions accumulated next to the edge of the brush.
The cumulative charge $\tilde{Q}(z)$ smoothly increases inside the brush and
passes through a maximum at the brush edge, $z\approx H$, and vanishes at $%
z\rightarrow \infty $ where the charge of the brush is fully neutralized by
the  conterions. This behavior of $\rho (z)$ and $\tilde{Q}(z)$
are consistent with the analytical theory predictions for the linear
elasticity regime.

At larger values of $\alpha =0.5$ and $\alpha =0.7$ non-linear elasticity
effects come into play and the onset of stratification takes place: At $%
z\approx 50$ corresponding to the boundary between inner in outer layers
(the position of this boundary $H_{1}$ can be estimated from the position of
the distal peak in the distribution of the branching points) $\rho (z)$
exhibits a sharp minimum (at $\alpha =0.7$ the value of $\rho (z)$ in the
minimum becomes negative) followed by a sharp maximum. This singularity of $%
\rho (z)$ gives rise to a small kink in $\tilde{Q}(z)$. The minimum in 
$\rho (z)$ at $z\approx H_{1}$ can be unambiguously attributed to a thin
cloud of counterions localized at the boundary between the layers, in
accordance with prediction of the approximate two-layer model. 


\begin{figure}[ht]
\centering{\includegraphics[scale=0.5]{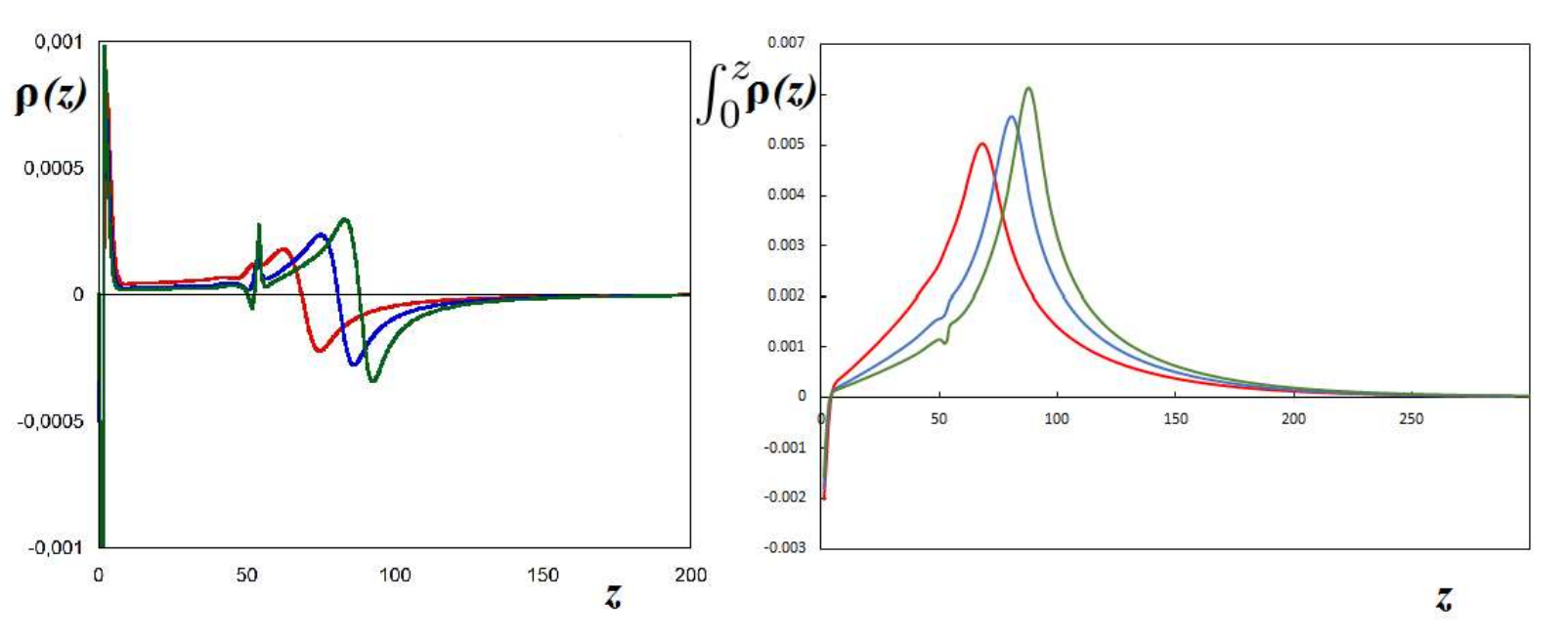}}   
\caption{Net local charge density $\protect\rho(z)$ (a) and its integral 
$\protect \tilde Q(z)=\int_{0}^z \protect\rho(z^{\prime})dz^{\prime}$ (b) plotted as a
function of teh distance from the grafting surface $z$. Red line corresponds to $\protect\alpha=0.3$, blue line
- $\protect\alpha=0.5$, green line - $\protect\alpha=0.7$.
Other parameters are $n=50,q=3, a^2/s=0.02$, salt volume fraction $c_s=10^{-5}$.}
\label{charge}
\end{figure}



\begin{figure}[ht]
\centering{\includegraphics[scale=0.5]{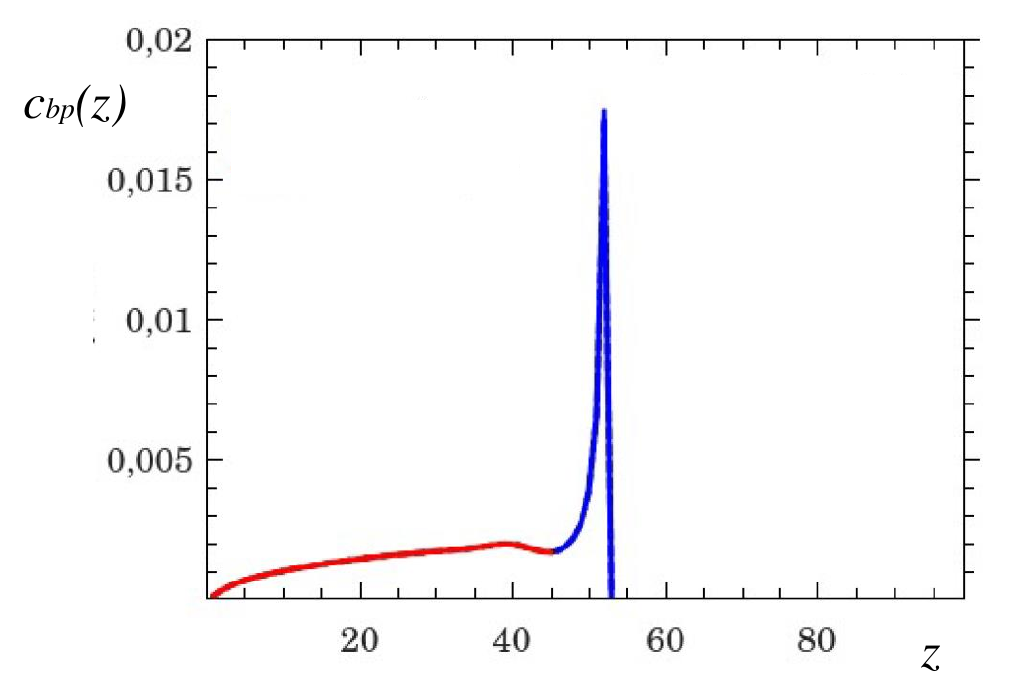}}   
\caption{ Branching points distribution for the brush of starlike
polyelectrolytes with $\protect\alpha=0.8, n=50,q=3, a^2/s=0.1, c_s=10^{-5}$.}
\label{phi_br}
\end{figure}

\begin{figure}[ht]
\centering{\includegraphics[scale=0.5]{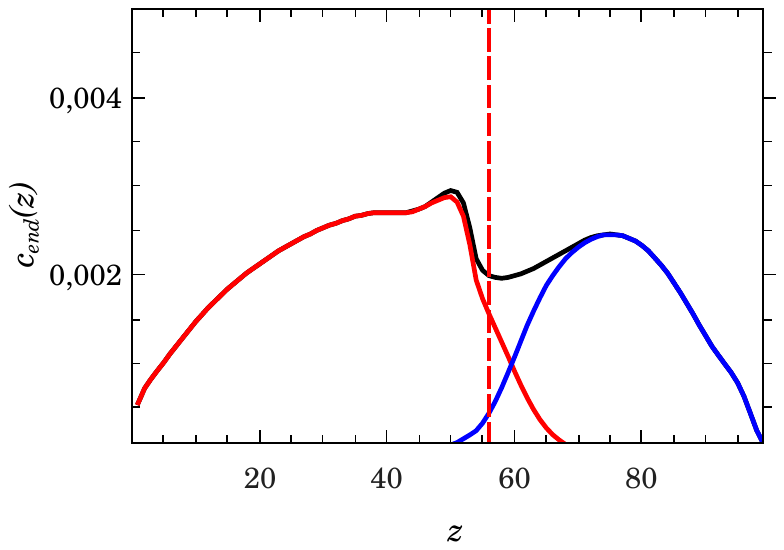}}   
\caption{The end-points distribution for the brush of starlike
polyelectrolytes with $\protect\alpha=0.8, n=50,q=3, a^2/s=0.1, c_s=10^{-5}$. Black line corresponds
to the cumulative distribution. The red and the blue lines corerspond to
partial distributions for the weakly and stronly stretched populations, respectively.}
\label{phi_end}
\end{figure}

\begin{figure}[ht]
\centering{\includegraphics[scale=0.5]{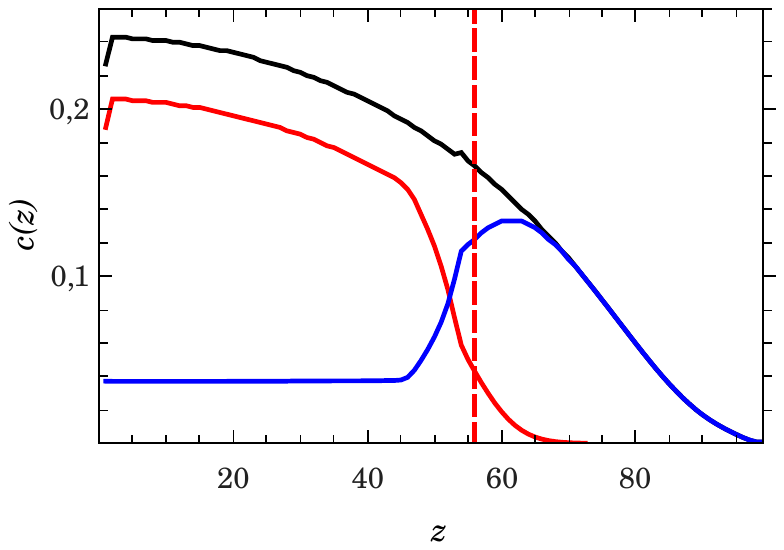}}   
\caption{ Polymer density distribution for the brush of starlike polyelectrolytes
with $\protect\alpha=0.8, n=50,q=3, a^2/s=0.1, c_s=10^{-5}$. Black line corresponds to the
cumulative density distribution. The red and the blue lines corerspond to
partial density distributions for the weakly and stronly stretched
populations, respectively.}
\label{phi_all}
\end{figure}

At even higher degree of ionization, $\alpha =0.8$, the brush acquires
well-developed two-layered (stratified) structure: We present the
distribution of branching points in Figure \ref{phi_br}, the end point
distribution for free arms in Figure \ref{phi_end} and the overall monomer
density distribution in Figure \ref{phi_all} .

The distribution of the branching points, Figure \ref{phi_br}, clearly shows
a sharp peak at limiting extension, $z\approx 50$, of the stems. This peak
corresponds to population of stars with strongly extended stems and less
extended free branches that form the upper layer of the brush. In the range
of $0\leq z\leq 45$ the branching points distribution is smooth and exhibits
a broad maximum. This part of the distribution corresponds to the population
of stars with weakly and moderately stretched stems. A minimum in the branching point
distribution observed at $z\approx 45$ can be approximately considered as
separating strongly weakly and strongly stretched populations.

The distribution of the free ends in Figure \ref{phi_end} exhibits two
pronounced maxima corresponding to weakly and strongly stretched populations
of stars. This cumulative distribution can be decomposed into two almost
non-overlapping partial distributions corresponding to weakly (with the
branching point position $z_{br}$ closer to the surface than the minimum at
the branching point distribution curve) and strongly stretched populations,
also indicated in the Figure \ref{phi_end} by red and blue curves,
respectively.

The overall monomer density distribution presented in Figure \ref{phi_all}
can be decomposed in a similar way. While for the weakly stretched population
the overall density profile is monotonously decreasing as a function of
distance from the surface $z$, for the strongly stretched population we
observe a plateau region at $0\leq z\leq 50$ corresponding to the fairly
uniformly extended stems whereas the major fraction of monomer units of the arms of
stars with strongly extended stems are distributed at $z\geq 50$.

\begin{figure}[ht]
\centering{\includegraphics[scale=0.5]{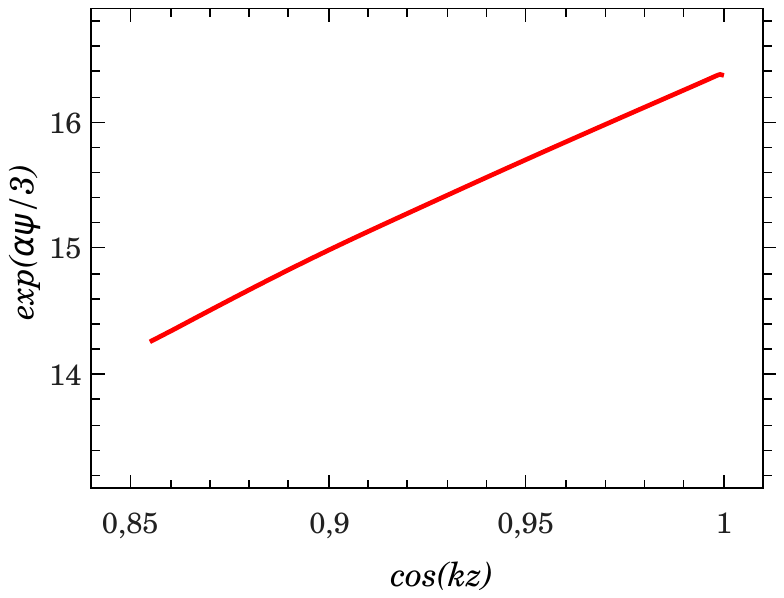}}   
\caption{ The electrostatic potential profile in the inner layer of the brush: $\exp \protect\alpha \protect\psi(z)/3$ plotted as a function of $\cos kz$
according to eq  \ref{psi_bilayer}, $\protect\alpha=0.8, n=50,q=3, a^2/s=0.1, c_s=10^{-5}$.}
\label{potfig}
\end{figure}

As soon as intra-brush segregation into two layered structure is apparent at
high degree of ionization, we check how well the distribution of local
charge density and the electrostatic potential follow predictions of the
analytical model. For this purpose we present in Figure \ref{potfig} the
dependence of $\exp [\alpha \psi (z)/3]$ plotted as a function of $\cos kz$
in the range of $0\leq z\leq H_{1}$, that is, inside the inner layer. In
good accordance with the first line of eq \ref{psi_bilayer} this dependence
is approximately linear.

\begin{figure}[ht]
\centering{\includegraphics[scale=0.5]{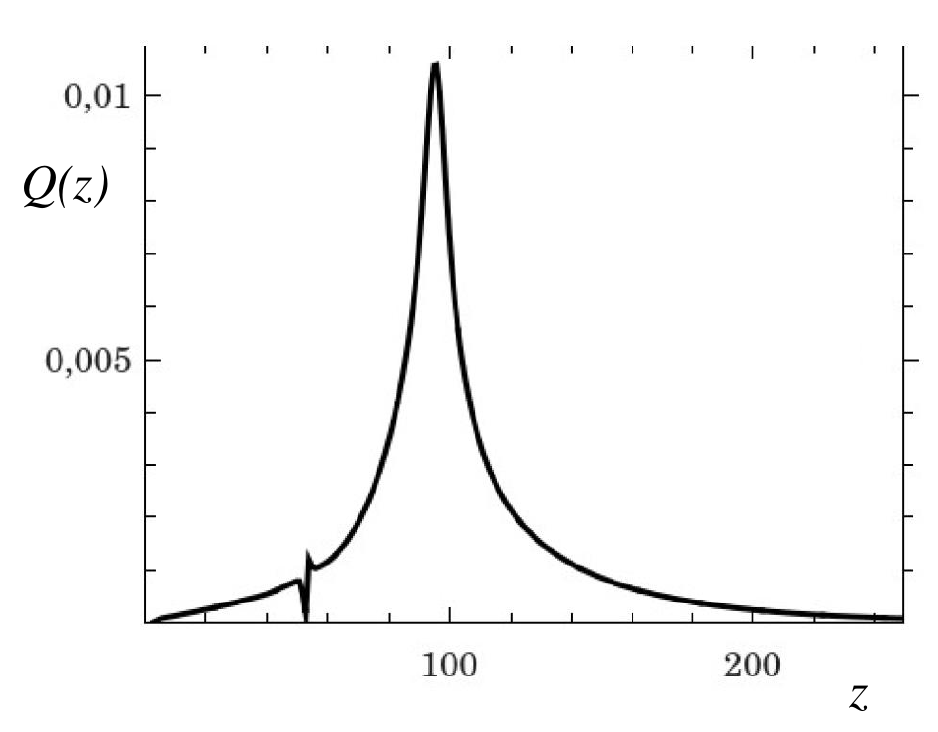}}   
\caption{Cumulative charge distribution, $\tilde Q(z)$, for $\protect\alpha=0.8, n=50,q=3, a^2/s=0.1, c_s=10^{-5}$}
\label{cum_charge}
\end{figure}

Finally, in Figure \ref{cum_charge} we present the plot of the cumulative
charge $\tilde{Q}(z)$ for $\alpha =0.8$. In accordance with predictions of
the analytical model the cumulative charge $\tilde{Q}(z)$ monotonously
increases as a function of $z$ in the range of $0\leq z\leq H_{1}$ and then sharply drops
to zero at $z=z_{min}\approx H_{1}$. This drop is due to the very thin cloud
of counterions localized at the boundary between the layers and neutralizing
the residual charge of the inner layer. The behavior of $\tilde{Q}(z)$ at $%
z\geq H_{1}$ is more complex than predicted by the analytical model: it
demonstrates a sharp peak at $z_{max}\geq z_{min}$ followed by subsequent
smooth growth. 

\section{Conclusions}

In this paper we investigated the effects of finite extensibility
(non-linear elasticity) in brushes formed by starlike polyions (dendrons of
the first generation) tethered by the end of one arm to a planar
solid-liquid interface. For that we used combination of the analytical
self-consistent field Poisson-Boltzmann theory and Scheutjens-Fleer
numerical approach.

Under conditions of low ionic strength of the solution and high degree of
ionization of the polyions, intermolecular electrostatic interactions lead
to  strong stretching of macromolecules up to the limit of extensibility of
the linear segments (tethers and free arms). Remarkably, in the case of
brushes formed by starlike polyelectrolytes, the limit of extensibility is
reached at lower degree of ionization compared to brushes of  linear chain
polyelectrolytes \cite{Lebedeva2017}.

The most interesting consequence of finite extensibility 
in brushes of polyelectrolyte stars is stratification related to 
disproportionation of stars into two populations of stronger and weaker stretched stars.
This stratification is unambiguously proven by our numerical calculations
which indicate bimodal distributions of the end segments of free arms and
of the branching points in the brush formed by strongly ionized stars. The
weaker stretched stars in the stratified brush are fully embedded into the
proximal to the surface layer. The branching points of the stars belonging
to the stronger stretched population are localized approximately at the
distance from the grafting surface corresponding to full extension of tethers
("stems"), while their free branches constitute the outer
layer. Similar stratification effect was predicted earlier for brushes
formed by non-ionic arm-tethered polymer stars in good solvent \cite%
{Polotsky10-1,Polotsky12-1,Zhulina2014}, but it took place at sufficiently
larger grafting densities.

We proposed an approximate two-layer analytical model of the stratified
brush formed by star-shaped polyelectrolytes on the basis of the
self-consistent  field Poisson-Boltzmann approximation with explicit account
of non-linear elasticity of the arms in the brush-forming stars. The
predictions of the analytical model were confronted to the results of the
numerical calculations based on the Scheutjens-Fleer method.

Remarkably, both the approximate analytical theory and the numerical model
point to the accumulation of a thin cloud of counterions (formation of
double electrical layer) near the boundary between inner and outer layers 
inside the stratified brush. 

The numerical calculations demonstrate a kink in the distributions of monomer
density (Figure \ref{phi_all}) and corresponding sharp maximum in cumulative
charge distribution (Figure \ref{cum_charge}) next to the position of the
counterion cloud at the boundary between the layers. We attribute this
singularity to localization of branching points of the dendrons of the
stronger stretched population close to the boundary between the layers and
discrete lattice implementation of the numerical self-consistent field
approach.

Hence, a combination of the simplified analytical and the approximation-free
numerical approaches enables to demonstrate and to rationalize stratified internal structure in the brush formed by branched
polyelectrolytes as well as to provide its comprehensive quantitative
description.

We remark that finite extensibility of spacers in brushes formed by ionically charged dendrons with larger 
number of generations results in stratification of the brush into multiple layers with corresponding multimodal 
distributions of the positions of the branching points  and terminal segments. However, the described effect of 
stratification is most pronounced and the two-layer structure is better distibguishable 
for the brush formed by the first generation dendrons studied here.


\section*{Acknowledgements}

This work was financially supported by Government of Russian Federation
(Grant 08-08) and by the European Union's Horizon 2020 research and innovation program under the Marie Sklodowska-Curie 
(grant agreement No 823883).


\end{document}